\documentclass{PoS}

\title{Radio emission from high-mass binaries with non-accreting pulsars}

\ShortTitle{Radio emission from binaries with pulsar}

\author{\speaker{Valent\'i Bosch-Ramon}\\
        Dublin Institute for Advanced Studies, Fitzwilliam Place 31, Dublin 2, Ireland\\
        E-mail: \email{valenti@cp.dias.ie}}

\abstract{Binary systems that harbor a non-accreting pulsar are efficient non-thermal emitters, from radio to gamma rays.
This broadband emission is thought to come from the region where the companion star and pulsar winds collide. A
paradigmatic example of this source type is PSR~B1259$-$63. Whereas the high-energy radiation probes the shock structure
at the binary scale, the radio emission is produced well outside the system, in regions where the shocked stellar and pulsar
winds are likely mixing due to hydrodynamical instabilities. Understanding the evolution of the shocked flow depends strongly
on a proper characterization of the low-energy radiation. We have performed numerical calculations of the radio emission
produced in a high-mass binary hosting a young pulsar. Adopting a prescription for the shocked flow in the system and the
non-thermal particle injection, we have generated synthetic radio emission maps that can be compared with observations.
Preliminary results suggest that the emitting bulk flow should be rather slow, with a pulsar wind magnetization at the shock
higher than in isolated pulsars.} 

\FullConference{25th Texas Symposium on Relativistic Astrophysics - TEXAS 2010\\
		December 06-10, 2010\\
		Heidelberg, Germany}

\begin{document}

\section{Introduction}\label{intro}

The X-ray binary PSR~B1259$-$63 is a long-period system, formed by a Be star and a young and powerful pulsar with spin-down luminosity $L_{\rm sd}\approx 8\times 10^{35}$~erg~s$^{-1}$
(\cite{man95}). Non-thermal emission in radio (e.g. \cite{joh05}), X-rays (e.g. \cite{com94,uch09,che09}) and gamma rays (e.g. \cite{aha05}) has been detected from this source. This
emission is thought to originate in the region where the star and pulsar winds collide (\cite{tav97}).  Therefore,  PSR~B1259$-$63 is not an accretion powered system, like many Be X-ray
binaries, since the pulsar wind ram pressure keeps the stellar wind beyond the neutron star gravitational capture radius (\cite{bon44}). The two-winds collision leads to two shocks, one in
the stellar wind, and another in the pulsar wind. The contact discontinuity between the two flows is located where the wind ram pressures are equal, at a minimum distance from the pulsar
$R_{\rm off}=\sqrt{\eta}\,R_{\rm orb}/(1+\sqrt{\eta})$, where $R_{\rm orb}$ is the star-pulsar separation distance, $\eta=L_{\rm dsd}/\dot{M}_*\,v_{\rm w}\,c$, and $\dot{M}_*$ and $v_{\rm
w}$ are the stellar mass loss rate and wind velocity, respectively (\cite{bog08}). Other three high-mass binaries in the Galaxy may harbor a non-accreting pulsar: LS~I~+61~303, LS~5039 and
HESS~J0632$+$057 (e.g. \cite{mar81,mar05,dub06,dha06,che06,sie08,hin09}), although their accreting nature cannot be discarded yet (e.g. see the discussion in \cite{bos09}; see also
\cite{rom07,mas09}). 

In binaries hosting a non-accreting pulsar, the X- and gamma rays are produced close or within the binary system, where the shocks are the strongest. X-rays are of likely synchrotron
origin, and gamma rays, of inverse Compton (IC) nature (e.g. \cite{kir99,kha07,sier08,tak09,ker11}; see however \cite{ner07}), and the cooling timescales of the electrons and positrons
emitting at these energies are probably very short. These particles are likely accelerated in the pulsar wind shock, more suitable for particle acceleration than the slower and less
energetic stellar wind shock. Accelerated particles are expected to follow a power-law of index $p$ ($\propto E^{-p}$) between a minimum ($E_{\rm min}$) and a maximum energy ($E_{\rm
max}$), cooling down while advected away in the shocked flow. At the scales of the shock, $\sim R_{\rm off}$, synchrotron and IC cooling dominate, but at farther distances energy losses are
dominated by adiabatic cooling (i.e. work). The velocity at which the flow leaves the shock region, and eventually the system, determines how much radiation is produced at places where
adiabatic cooling is dominant. This is so because the non-thermal power scales as $\propto 1/t_{\rm ad}\sim R/v_{\rm esc}$, where $R$ and $v_{\rm esc}$ are the characteristic size of the
flow and the shock escape velocity, respectively. The emission coming from the outskirts of the binary and beyond can be better studied in radio, and provides information on the material
flowing away from the system. An important factor affecting the colliding wind region is the development of instabilities, which can mix the shocked pulsar wind with the much denser and
colder stellar wind. 

High-resolution VLBI studies can be of very much help characterizing the shocked flows. Remarkably, extended radio emission has been recently detected from PSR~B1259$-$63 (\cite{mol11}; see
also \cite{dha06} and \cite{rib08} regarding the candidates LS~I~+61~303 and LS~5039). To fully profit from these observations, however, detailed radiation modeling is required. From the
radio emission from binaries hosting a young pulsar, one can derive important information of the shocked medium, like its velocity and magnetic field, both to be affected by mixing. The
emitting particles themselves can also be studied, since the value of $E_{\rm min}$ affects strongly the number of the radio emitting particles. It is useful to compute maps of the radio
emitter, to compare with observations. A study of the radio emission appearance was already done in \cite{dub06}, although the radio emitter was treated there as 1-dimensional, with
point-like injection. In this work, we present calculations done adopting a prescription for the shocked star-pulsar wind structure that accounts for the 3-dimensional extension of the
particle injection and the radio emitter. The results are preliminary, but they can already shed light on the properties of the radio emitting flow in binaries hosting a non-accreting
pulsar.  In Figure~1 we show a sketch of the considered problem.

\begin{figure}[]
\begin{center}
%\resizebox{\hsize}{!}{\includegraphics{sk.eps}}
\includegraphics[width=0.5\textwidth]{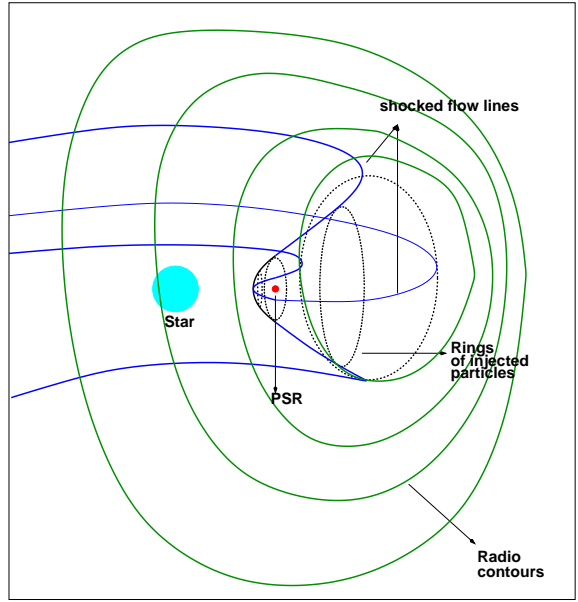}
\caption{Sketch of the scenario and the model presented here. The pulsar and the stellar wind collide forming two shocks
separated by a contact discontinuity. The shocked material flows away and gets spiral-shaped by the orbital motion of the
system. Electrons and positrons are injected in the interaction surface, in the form of different rings, 
at different distances from the 
pulsar. These particles flow with the shocked material producing radio to X-rays through synchrotron, and X-rays to gamma
rays through IC scattering. Here we concentrate in the radio emission. Overlaid on the system and interaction region, the
contours of the source radio image are shown.}
\label{p1}
\end{center}
\end{figure}

\section{The model}

We model the radio emitter injecting particles, electrons and positrons, as rings. These rings are distributed along a paraboloid that is connected to a cone beyond a certain distance. The
flux of energy at the shock, going to non-thermal particles, is computed from $\sin(\theta)\,L_{\rm sd}/4\pi\,l_{\rm p0}^2c$, where $\theta$ is the angle between the shock surface and the
pulsar radial direction, and $l_{\rm p0}$ is the distance from a particular injection point to the pulsar (or characteristic injection region size). The paraboloid is initially defined as
$r=R_{\rm orb}\sqrt{y-d_{\rm off}}$, where $r$ is the radius of the paraboloid, $y$ the distance to the star from the paraboloid axis, and $d_{\rm off}=R_{\rm orb}-R_{\rm off}$. The final
opening angle of the cone is: $\approx 0.5\,(4-\eta^{2/5})\,\eta^{1/3}$, which corresponds in reality to the opening angle of the contact discontinuity between the two winds \cite{bog08},
simplified here as the whole shocked structure.  Particles are advected away along the parabolic/conical surface in the direction opposite to the star. In fact, these rings of particles are
injected, with time, in different locations following the pulsar orbital motion around the companion. Once injected, rings are assumed to follow ballistic motion, and thus the shape of the
whole structure at large scales tends to form a spiral. Note that ballistic motion is just a rough approximation; the shocked stellar wind exerts a Coriolis force on the pulsar wind,
bending it even farther in the direction opposite to the pulsar orbital motion.  

We do not adopt here the Kennel \& Coroniti solution (\cite{ken84a}) for the postshock flow, as it was done in \cite{dub06}. The assumption of a flow being like the one treated in
\cite{ken84a}, of spherical nature, is not realistic here since the flow evolution is strongly affected by the inhomogeneity and anisotropy of the pulsar environment (i.e. the stellar
wind). This is clearly seen in the results of \cite{bog08}. However, instabilities were not considered in that work, and simulations including instability development have not been
performed yet. For this reason, some parameters are treated here phenomenologically. The flow velocity is derived defining a velocity of the shocked flow along its trajectory, parabolic
first, straight later on. The velocity is taken to be $c/3$ at the injection locations, although this is only strictly true at the regions in which the pulsar shock is roughly
perpendicular. At distances bigger than the injection region ($>l_{\rm p0}$), we have assumed that the flow speed decreases exponentially down to some intermediate velocity $v_{\rm
w}<v_{\rm esc}<c/3$ (typically $v_{\rm w}\sim (1-2)\times 10^8$~cm~s$^{-1}$). This decrease in the bulk flow velocity relates to mixing due to hydrodynamical instabilities\footnote{As shown
in \cite{bog08}, a hydrodynamical approximation for the flow seems appropriate.} in the contact discontinuity, most likely of Kelvin-Helmholtz nature. Kelvin-Helmholtz instabilities would
start to grow at scales of the order of those of the shock region, thus mainly affecting the radio emitting region. Eventually, the final turbulent and mixed flow velocity should be close
to the one obtained assuming momentum conservation, due to significant kinetic to internal/turbulent energy conversion. In this work, $v_{\rm esc}$ is fixed to $10^9$~cm~s$^{-1}$ for
simplicity. Adiabatic losses have been computed as given in Sect.~\ref{intro}. To compute synchrotron losses, we have adopted a magnetic field $B_0\approx 3\sqrt{\sigma\,2\,L_{\rm
sd}/l_{\rm lp0}^2\,c}$ (\cite{ken84b}), and $B$ decreases farther in the shocked flow as $1/l_{\rm p}$. Turbulent mixing should affect $B$, but we have not accounted for it at this stage.
The value of $E_{\rm min}$ has been fixed to $\sim 0.1\Gamma_{\rm p}\,m_{\rm e}\,c^2$, where $\Gamma_{\rm p}$ is the Lorentz factor of the particles in the pulsar wind (\cite{ken84b}),
fixed by us to $10^5$. The value of $E_{\rm max}$ has been set to 1~TeV. The spin-down luminosity of the pulsar has been taken $L_{\rm sd}=10^{36}$~erg~s$^{-1}$. Regarding the system
properties, the orbit is circular, with $R_{\rm orb}=3\times 10^{12}$~cm and an inclination angle $i=45^\circ$. The inferior conjunction of the compact object corresponds to phase 0.5. The
star luminosity has been fixed to $5\times 10^{38}$~erg~s$^{-1}$, and the temperature to $3\times 10^4$~K. The stellar mass loss rate is $\dot{M}_*=10^{-6}\,M_{\odot}$/yr, and $v_{\rm
w}=2\times 10^8$~cm. The IC losses are derived from the stellar photon energy density (see \cite{bos09} and references therein). Densities are low enough to neglect ionization/coulombian
losses and relativistic Bremsstrahlung.

The radio emission has been calculated from the evolved particle populations after propagating from their injection point, at $l_{\rm p0}$. Synchrotron self-absorption, and free-free
absorption in the stellar wind (ionization fraction of 0.1), have been taken into account. The impact of these absorption processes is rather small provided that most of the radio emission
comes from a fairly large region, well outside the binary.

\section{Results and discussion} 

\begin{figure}[]
\begin{center}
\includegraphics[width=0.7\textwidth]{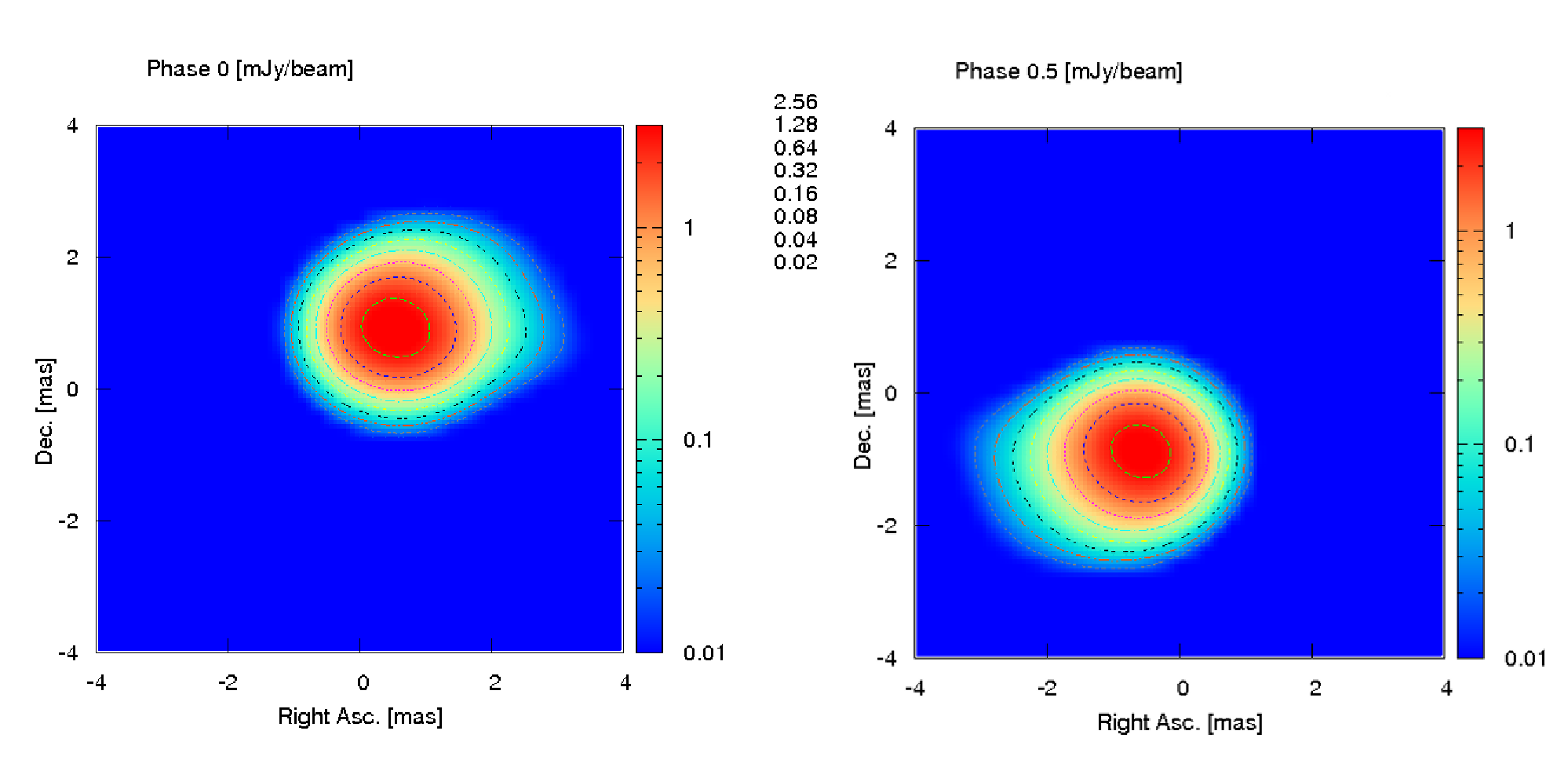}
\includegraphics[width=0.7\textwidth]{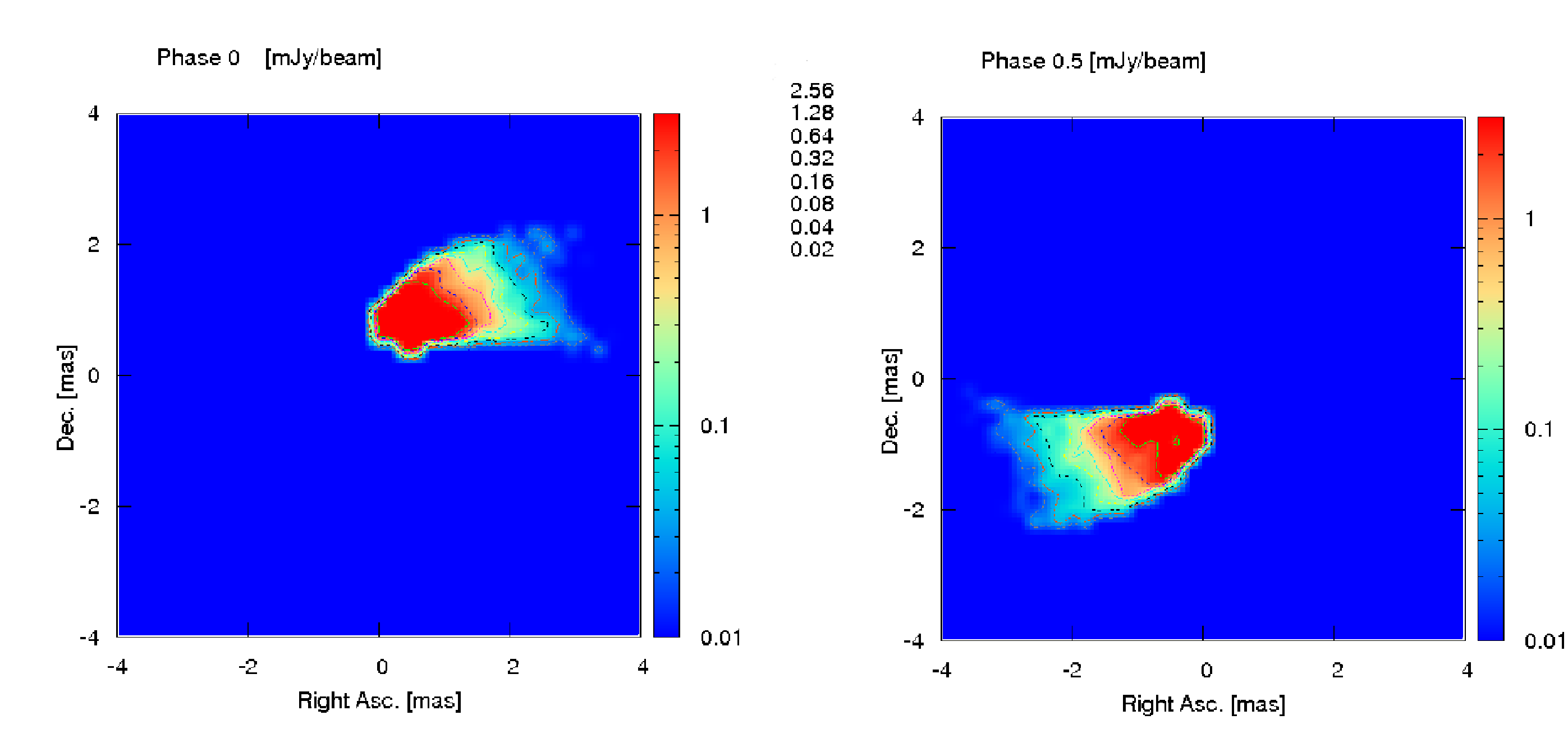}
\caption{Top: Simulated radio map at 5~GHz, smoothed with a gaussian of FWHM $\approx 1$~mas, 
for the generic case studied here and phases 0 and 0.5. Intensity units are mJy per beam. Axis units are in mas.
Bottom: The same as in the top of the figure, but without gaussian smoothening.}
\label{p1}
\end{center}
\end{figure}

In Figure~2, top, 5~GHz radio maps are computed for a generic binary system and two different orbital phases. The maps have been obtained smoothening the computed flux-per-beam values with
a 1~milliarcsecond (mas) FWHM Gaussian, reproducing the effect of a radio interferometer with a 1~mas beam. In Figure~2, bottom, the same maps without Gaussian smoothening are also shown.
From Fig.~2, it is clear that the overall picture differs strongly from a point-like source. The emitter spiral shape is hinted in Fig.~2, top, and much more clear in Fig.~2, bottom. The
center of gravity is displaced by few mas between phases 0 and 0.5, and the total flux in both phases is $\approx 13$~mJy. Despite the flow prescriptions are quite distinct, and that we
have accounted for the 3-dimensional extension of particle injection and the emitter, our results and those of \cite{dub06} are qualitatively similar. 

The adopted value for $\sigma$, 0.1, may appear rather high. In the case of Crab, the parameter $\sigma$ is much lower (\cite{ken84a}). A higher $\sigma$ in our case is however plausible if
the transfer of energy, from the Pointing flux to matter in the pulsar wind, is a process extended in space. The conversion may even take place in the termination shock itself
(\cite{pet07}). Since $R_{\rm off}$ in a close binary system is much smaller than in Crab, the $\sigma$-value may well be significantly higher in the former case. Another possibility would 
be an effective increase (or decrease) of $B$ through the mixing of the shocked stellar and pulsar winds, as the radio emitting plasma gets entangled with the stellar wind magnetic field.

Concerning $v_{\rm esc}$, simulations predict values close to $c$ due to the acceleration of the shocked flow because of strong pressure gradients \cite{bog08}. Under such conditions, as
shown in \cite{kha08}, adiabatic cooling timescales can be indeed very short. However, instabilities and mixing at scales significantly larger than $R_{\rm off}$ should slow down the flow.
This has been included here and allow radio fluxes to reach significant values. Otherwise, the emission would be lower by more than one order of magnitude, i.e. radio fluxes $<1$~mJy. 

Nowadays, radio observations start to allow detailed comparisons with models. In this paper, we present preliminary results of such a comparisons in the context of a semi-phenomenological
work, based on current state of our knowledge on the flow structure. A more detailed study of the properties of the radio emitter, as well as applications to specific objects, will be
presented elsewhere. We remark that, to better characterize the shocked wind evolution, farther numerical magnetohydrodynamics simulations are required, with a detailed study of the
instability development. Pulsar wind physics is also a very important but open issue, central for wind propagation, particle acceleration at the wind termination, and postshock region
conditions.

\acknowledgments
V.B-R. want to thank Dmitry Khangulyan for very helpful discussions on the topic treated here. 
The research leading to these results has received funding from the European Union
Seventh Framework Program (FP7/2007-2013) under grant agreement
PIEF-GA-2009-252463.
V.B.-R. acknowledges support by the Spanish Ministerio de Ciencia e 
Innovaci\'on (MICINN) under grants AYA2010-21782-C03-01 and 
FPA2010-22056-C06-02.

\end{document}